# Photothermal Absorption Spectroscopy of Individual Semiconductor Nanocrystals

*Stéphane Berciaud, Laurent Cognet & Brahim Lounis**

Centre de Physique Moléculaire Optique et Hertzienne, CNRS (UMR 5798) et Université Bordeaux I,

351, cours de la Libération, 33405 Talence Cedex, France

*Corresponding author. E-mail: b.lounis@cpmoh.u-bordeaux1.fr

ABSTRACT: Photothermal Heterodyne Detection is used to record the first room temperature absorption spectra of single CdSe/ZnS semiconductor nanocrystals. These spectra are recorded in the high cw excitation regime and the observed bands are assigned to transitions involving biexciton and trion states. Comparison with the single nanocrystals photoluminescence spectra leads to the measurement of spectral Stokes shifts free from ensemble averaging.

Core shell CdSe/ZnS semiconductor nanocrystals exhibit unique size-dependent optical properties[1] which make them ideal candidates for applications in various fields such as the design of new optical devices[1-3] or bio-labeling[4]. For full exploitation, understanding of their optical properties is however crucial. Ensemble photoluminescence and linear absorption measurements provided a comprehensive description of the excitonic levels structure of CdSe/ZnS nanocrystals[5] and allowed for a precise assignment of the optical transitions[6]. Individual nanocrystals are now commonly detected using standard fluorescence microscopes with low excitation intensities due to the high radiative quantum yield of their band edge exciton state X ($1S_e$, $1S_{h,3/2}$) (see reference 6 for spectroscopic notations). Their luminescence displays a blinking[7] behavior with intensity dependent on-times distributions and strong photon anti-bunching [8, 9].

CdSe/ZnS nanocrystals have relatively high absorption cross-sections[8, 10] (typically $\sim 10^{-15} cm^2$) and when excited with sufficiently high intensities, excitons are created at average rates $N_{abs}$ significantly higher than their radiative recombination rates $\Gamma_{rad}$[8, 11]. In this regime, efficient non-radiative Auger recombinations of the prepared multiexcitons take place[12]. In CdSe/ZnS nanocrystals, these recombinations occur in the picoseconds range as measured in time resolved experiments on ensemble samples[13]. Individual nanocrystals could thus be detected through their absorption using the Photothermal Heterodyne method[14].

Although recording the luminescence spectra from single nanocrystals has become routine, a measurement of the absorption spectra of individual nanocrystals has however never been reported due to a lack of sensitive enough methods. In this letter, we present the first room temperature photothermal absorption spectra of individual nanocrystals.

In the photothermal techniques[14, 15] the signal is proportional to the amount of laser power absorbed by a single nanoparticle converted into heat by non-radiative relaxation processes. This allowed to detect individual tiny metal nanoparticles and to perform spectroscopy of their Plasmon resonance[16]. For nanocrystals the signal has its origin from two possible non-radiative relaxation pathways. First, at the high cw excitation rates required in the photothermal method (typically $N_{abs} \sim 1 ns^{-1}$), biexcitons are created in the nanocrystal and the $(1S_e, 1S_{h,3/2})(1S_e, 1S_{h,3/2})$ biexciton state (XX) is prepared. A rapid recombination of one of the two excitons through Auger processes occurs ($\Gamma_{XX} > 20 ns^{-1}$ for 2nm nanocrystals[13]) and is followed by a non radiative relaxation of the remaining exciton to the band-edge state X. Since $N_{abs}$ is much higher than $\Gamma_{rad}$ ($\sim 1/20 ns^{-1}$), multiexciton creation efficiently prevents radiative recombination of the monoexcitons. Thus, the first contribution to the signal arises from continuous cycling of the nanocrystal between the band-edge X and biexciton XX states (see Fig. 1). The second mechanisms contributing to the signal come into play when the nanocrystal is in a charged state. It is known that continuous excitation of a CdSe nanocrystal can lead to a spontaneous or Auger type ejection of charges to the shell or to the local environment of the nanocrystal[7, 17, 18]. In this case, absorption leads to the formation of trions and the photothermal signal arises from fast Auger non-radiative recombination of X* state to the ground charge state 0* ($\Gamma_{X*} \sim 20 ns^{-1}$)[19]. In addition, with our experimental parameters, the probability to prepare triexcitons from XX or charged biexcitons from X* is very low ($N_{abs} << \Gamma_{XX}, \Gamma_{X*}$).

The absorption spectroscopy setup (Fig. 2) using Photothermal Heterodyne signals was described elsewhere[16]. In short, it consists in a heating beam (tunable cw dye laser, Coherent 599, tuning range 530–590 nm corresponding to photon energies of 2.10-2.34 eV) overlaid with a probe beam (HeNe laser, 633nm wavelength, well below the band-edge absorption of the nanocrystals). The heating beam is circularly polarized with an intensity of $\sim 1 MW/cm^2$ corresponding to absorption rates of $N_{abs} \sim 1 ns^{-1}$. Both beams are focused on the sample using a high NA objective. A fluorescence confocal microscopy scheme is also implemented on this setup. It is used to image individual luminescent nanocrystals and to record their luminescence spectra. For these measurements, the heating laser is used for the excitation with an attenuated intensity ($\sim 1 kW/cm^2$) and a fixed energy (2.30 eV). The collected luminescence photons are split towards an avalanche photodiode and a spectrometer. The samples were prepared by spin-coating dilute solutions of nanocrystals (average radius: $2.0\,nm$, peak emission at $\sim 2.10\,eV$, dissolved in 1% mass polyvinyl-alcohol aqueous solution) onto clean microscope coverslips. The dilution and spinning rates were chosen such that the final density of nanoparticles in the samples is less than $1\,\mu m^{-2}$. A drop of viscous silicon oil was added on top of the samples to ensure homogeneity of heat diffusion.

Individual fluorescent nanocrystals were first located by recording a $20\times 20\,\mu m^2$ fluorescence image of the sample (Fig.3 (a)). For each spot, a luminescence time trace and an emission spectrum were recorded simultaneously. The blinking behavior, a signature of single nanocrystal emission, is clearly visible in the time trace shown on Fig.3 (c). We then recorded a photothermal image of the same sample area (Fig.3 (b)). Cleary the luminescence and photothermal images correlate well ($>80\%$ of the luminescence spots correlate with a photothermal spot), proving that the spots in the photothermal image stem from individual nanocrystals. Contrary to luminescence, the photothermal signals do not show any blinking behavior (Fig.3 (d)). They remain stable during time scales larger than those necessary to record an absorption spectrum.

As expected, we observe a large heterogeneity of luminescence spots heights due to diversity in nanocrystals structures, surface states and orientations with the respect to the polarization of the excitation light. As can be seen on Fig.3 (b) the photothermal signal levels recorded from individual nanocrystals are also heterogeneous. However, we cannot establish a clear correlation between the heterogeneities of the luminescence and photothermal signal levels. We also notice that some non-luminescent nanocrystals (which do not appear in Fig.3 (a)) are observed on the photothermal image. Furthermore, the photothermal signal scales linearly with the heating intensity (Fig.3 (e)) as the transitions involved in the signal are not saturated ($N_{abs} << \Gamma_{XX}, \Gamma_{X^*}$).

We acquired the photothermal absorption spectra for the nanocrystals which colocalize in the luminescence and photothermal images. Figure 4 (top) presents the results obtained from the individual nanocrystal highlighted on Figure 3. For the absorption spectrum, the laser frequency was scanned with steps of 4 meV and $200\,ms$ integration times per point. The signals are corrected for the wavelength dependence of the diffraction-limited laser spot size and the heating power at the sample was measured during the acquisitions for normalization.

Contrary to the ensemble luminescence spectrum, which has an inhomogeneously broadened Gaussian shape (open circles of Fig. 4 (bottom) with peak energy at $2.100\,eV$ and full-width-at-half-maximum (FWHM) of $100\,meV$), the luminescence spectrum displays a narrower homogeneous Lorentzian shape peaked at $2.120\,eV$ and a FWHM of $65\,meV$ (open circles on Fig. 4 (top)). The absorption spectrum of the same nanocrystal (measured with the photothermal method) and that of an ensemble of nanocrystals (measured with a spectrophotometer) are also shown on Fig. 4 (solid lines). Both of them contain a first absorption peak at respectively 2.160 $eV$ and 2.170 $eV$, blue shifted from the emission lines. Similarly to the emission spectrum, the width of the single nanocrystal absorption peak is narrower than that of the ensemble. The two peaks of the ensemble absorption spectrum, recorded in the weak excitation regime, are assigned to the band-edge X state and to the higher energy ($1S_e$, $2S_{h,3/2}$) state[6, 20] (Fig. 4 (bottom, solid line)). The non-resonant Stokes shift $\Delta E_S$ is defined as the energy difference between the first absorption peak and the luminescence maximum energy. We found $\Delta E_S$ ~ 70 meV in agreement with previous ensemble studies[21],

The photothermal absorption spectrum presented on Fig. 4(top, solid line) exhibits one clear band and a rising feature at higher energies which is not fully scanned due to insufficient laser tuning range. We assign the low energy band to the two unresolved $X \rightarrow XX$ and $0^* \rightarrow X^*$ transitions. Indeed, Coulomb interactions induce comparable red-shifts of the transitions[22] with respect to $E_X$, the energy of the $0 \rightarrow X$ transition (see Fig. 1). The rising feature in the spectrum may involve $X \rightarrow XX'$ and $0 \rightarrow X'^*$ transitions where $X'$ denotes a high energy exciton states (such as ($1S_e$, $2S_{h,3/2}$) or ($1P_e$, $1P_{h,3/2}$)).

We performed the same measurements for 23 individual nanocrystals and observed qualitatively comparable spectra. For each of them, we extracted the energy separation $\Delta E$ between the position of the low energy absorption band and the luminescence peak. The histogram of the values of $\Delta E$ is shown on Fig.4 (inset). It is well fitted by a Gaussian distribution with a FWHM of $12\,meV$ and a mean value of ~50 *meV* smaller by ~20 *meV* than $\Delta E_S$. This deviation is in agreement with the values of biexciton and trion binding energies deduced from femtosecond photoluminescence up-conversion[23], transient photoluminescence[24] studies (~20-30 *meV* for 2nm nanocrystal) as well as from the emission properties of single colloidal nanocrystals[25] or self-assembled quantum dots[22] in charged states.

In conclusion we have demonstrated the applicability of the Photothermal Heterodyne method to room temperature detection and absorption spectroscopy of individual CdSe/ZnS semiconductor nanocrystals in the multi-excitonic regime. We foresee promising spectroscopic studies of nanocrystals using this method with samples at cryogenic temperatures.

ACKNOWLEDGMENTS. We wish to thank O. Labeau, P. Tamarat and G. Blab for helpful discussions and A. L. Efros for the reading of the manuscript. This research was funded by CNRS (ACI Nanoscience and DRAB), Région Aquitaine and the French Ministry for Education and Research (MENRT).

FIGURE CAPTIONS

**Figure 1.** Energy diagram of the excitonic states involved in the photothermal signal from nanocrystals in the neutral and charged state. For notations see details in the text.

**Figure 2.** Schematic of the experimental setup.

**Figure 3**. (a) Comparison of photoluminescence (b) and Photothermal Heterodyne images of the same area $20 \times 20\,\mu m^2$ of a sample containing CdSe/ZnS nanocrystals. The integration time per point was 10 ms. Scale bar is $2\,\mu m$. Photoluminescence (c) and photothermal (d) signals recorded over time for the individual nanocrystal highlighted on (a) and (b). The integration time per point was 100 ms. (e) Photothermal signal obtained from an individual nanocrystal (circles) as a function of the heating intensity. The heating photon energy was $2.30\,eV$. The data are adjusted by a linear fit (solid line).

**Figure 4.** Top: Photoluminescence (open circles) and photothermal absorption (solid lines) spectra recorded for the individual nanocrystal highlighted in Fig. 3. For clarity a smoothing of the absorption spectra is shown. Bottom: Photoluminescence (open circles) and linear absorption spectra (solid line) of an ensemble of nanocrystals. Inset: Histogram of the values of ΔE obtained from photoluminescence and photothermal absorption spectra of 23 individual nanocrystals. The distribution is fitted by a Gaussian curve (black line).

REFERENCES

1. Klimov, V. I., Semiconductor and Metal Nanocrystals: Synthesis, Electronic and Optical Properties. Marcel Dekker: New York, 2003.

2. Colvin, V. L.; Schlamp, M. C.; Alivisatos, A. P., Light-emitting diodes made from cadmium selenide nanocrystals and a semiconducting polymer. Nature **1994,** 370, 354 - 357.

3. Dabbousi, B. O.; Bawendi, M. G.; Onitsuka, O.; Rubner, M. F., Electroluminescence from CdSe quantum-dot/polymer composites. Applied Physics Letters **1995,** 66, 1316-18.

4. Chan, W. C.; Maxwell, D. J.; Gao, X.; Bailey, R. E.; Han, M.; Nie, S., Luminescent quantum dots for multiplexed biological detection and imaging. Curr Opin Biotechnol **2002,** 13, (1), 40-6.

5. Efros, A. L.; Rosen, M.; Kuno, M.; Nirmal, M.; Norris, D. J.; Bawendi, M., Band-edge exciton in quantum dots of semiconductors with a degenerate valence band: Dark and bright exciton states. Physical Review. B. Condensed Matter. **1996,** 54, (7), 4843-4856.

6. Norris, D. J.; Bawendi, M. G., Measurement and assignment of the size-dependent optical spectrum in CdSe quantum dots. Physical Review. B. Condensed Matter. **1996,** 53, (24), 16338-16346.

7. Nirmal, M.; Dabbousi, B. O.; Bawendi, M. G.; Macklin, J. J.; Trautman, J. K.; Harris, T. D.; Brus, L. E., Fluorescence intermittency in single cadmium selenide nanocrystals. Nature **1996,** 383, 802-804.

8. Lounis, B.; Bechtel, H. A.; Gerion, D.; Alivisatos, A. P.; Moerner, W. E., Photon antibunching in single CdSe/ZnS quantum dot fluorescence. Chem.Phys.Lett. **2000,** 329, 399.

9. Michler, P.; Imamoglu, A.; Mason, M. D.; Carson, P. J.; Strouse, G. F.; Buratto, S. K., Quantum correlation among photons from a single quantum dot at room temperature. Nature **2000,** 406, (6799), 968-70.

10. Leatherdale, C. A.; Woo, W.-K.; Mikulec, F. V.; Bawendi, M. G., On the absorption cross section of CdSe nanocrystal quantum dots. J. Phys. Chem. B **2002,** 106, 7619.

11. Schlegel, G.; Bohnenberger, J.; Potapova, I.; Mews, A., Fluorescence decay time of single semiconductor nanocrystals. Phys Rev Lett **2002,** 88, (13), 137401.

12. Efros, A. L.; Lockwood, D. J.; Tsybeskov, L., Semiconductor Nanocrystals. Kluwer Academic / Plenum Publishers: New-York, 2003.

13. Klimov, V. I.; Mikhailovsky, A. A.; McBranch, D. W.; Leatherdale, C. A.; Bawendi, M. G., Quantization of multiparticle auger rates in semiconductor quantum dots. Science **2000,** 287, (5455), 1011-3.

14. Berciaud, S.; Cognet, L.; Blab, G. A.; Lounis, B., Photothermal Heterodyne Imaging of Individual Nonfluorescent Nanoclusters and Nanocrystals. Phys Rev Lett **2004,** 93, (25), 257402.

15. Boyer, D.; Tamarat, P.; Maali, A.; Lounis, B.; Orrit, M., Photothermal Imaging of Nanometer-Sized Metal Particles Among Scatterers. Science **2002,** 297, (5584), 1160-1163.

16. Berciaud, S.; Cognet, L.; Tamarat, P.; Lounis, B., Observation of Intinsic size effects in the optical response of individual gold nanoparticles. Nano Letters **2005,** 5, (3), 515-518.

17. Efros, A. L.; Rosen, M., Random Telegraph Signal in the Photoluminescence Intensity of a Single Quantum Dot. Phys Rev Lett **1997,** 78, 1110-1113.

18. Krauss, T. D.; O'Brien, S.; Brus, L. E., Charge and Photoionization Properties of Single Semiconductor Nanocrystals. J. Phys. Chem. B **2001,** 105, 1725-1733.

19. Wang, L. W.; Califano, M.; Zunger, A.; Franceschetti, A., Pseudopotential theory of Auger processes in CdSe quantum dots. Phys Rev Lett **2003,** 91, (5), 056404.

20. Ekimov, A. I.; Hache, F.; Schanne-Klein, M. C.; Ricard, D.; Flytzanis, C.; Kudryavtsev, I. A.;

Yazeva, T. V.; Rodina, A. V.; Efros, A. L., Absorption and intensity-dependent photoluminescence measurements on CdSe quantum dots: assignment of the first electronic transitions. JOSA B **1993,** 10, (1), 100-107.

21. Kuno, M.; Lee, J. K.; Dabbousi, B. O.; Mikulec, F. V.; Bawendi, M. G., The band edge luminescence of surface modified CdSe nanocrytallites: Probing the luminescing state. J. Chem. Phys. **1997,** 106, (23), 9869.

22. Patton, B.; Langbein, W.; Woggon, U., Trion, biexciton, and exciton dynamics in single self-assembled CdSe quantum dots. Phys. Rev. B **2003,** 68, 125316.

23. Achermann, M.; Hollingsworth, J. A.; Klimov, V. I., Multiexcitons confined within a subexcitonic volume: Spectroscopic and dynamical signatures of neutral and charged biexcitons in ultrasmall semiconductor nanocrystals. Phys. Rev. B **2003,** 68, 245302.

24. Caruge, J. M.; Chan, Y.; Sundar, V.; Eisler, H. J.; Bawendi, M. G., Transient photoluminescence and simultaneous amplified spontaneous emission from multiexciton states in CdSe quantum dots. Phys. Rev. B **2004,** 70, 085316.

25. Shimizu, K. T.; Woo, W. K.; Fisher, B. R.; Eisler, H. J.; Bawendi, M. G., Surface-enhanced emission from single semiconductor nanocrystals. Phys Rev Lett **2002,** 89, (11), 117401.

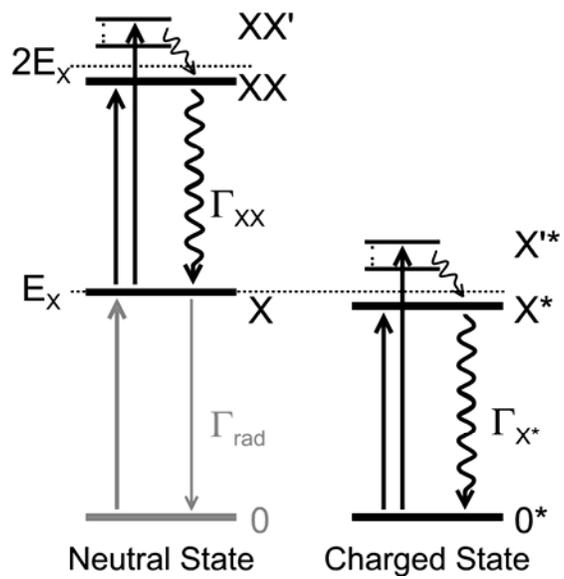

Figure 1

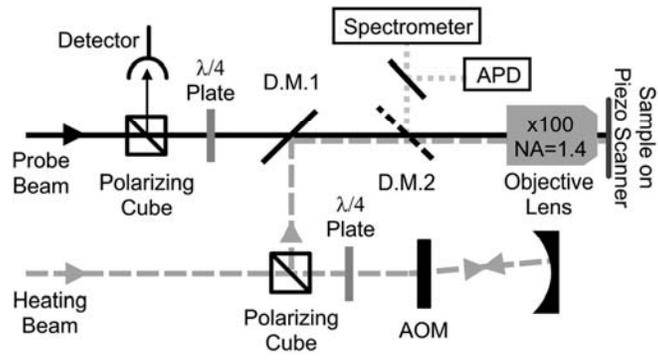

Figure 2

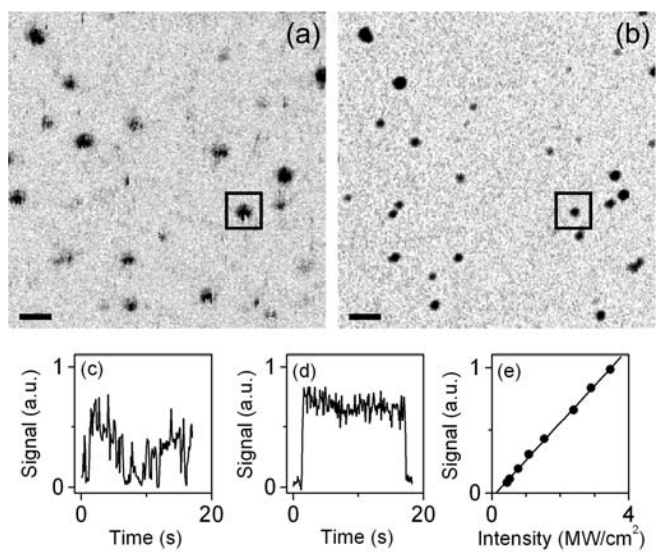

Figure 3

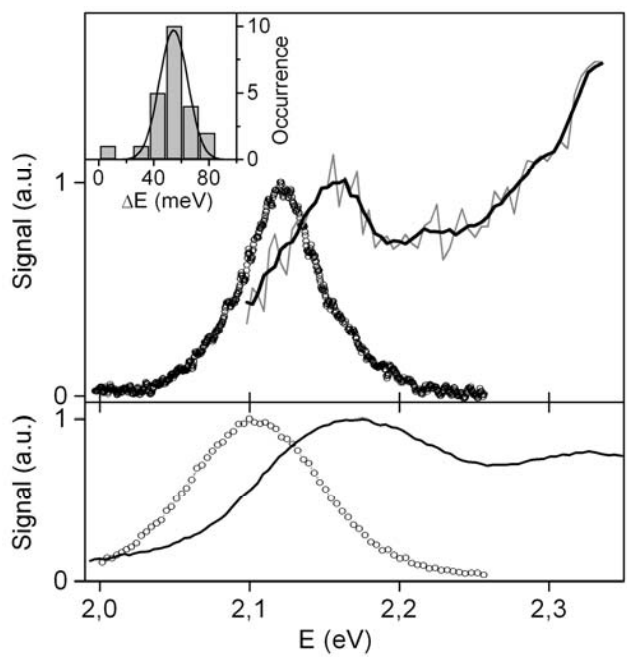

Figure 4